\documentclass[10pt,twocolumn,letterpaper]{article}

\usepackage{cvpr}              

\usepackage{newfloat}
\usepackage{listings}
\usepackage{booktabs}
\usepackage{multirow}
\usepackage{amsmath}
\usepackage{amssymb}
\usepackage{times}
\usepackage{helvet}
\usepackage{courier}
\usepackage{xcolor}

\definecolor{cvprblue}{rgb}{0.21,0.49,0.74}
\usepackage[pagebackref,breaklinks,colorlinks,allcolors=cvprblue]{hyperref}

\def\paperID{43932} 
\def\confName{CVPR}
\def\confYear{2026}

\title{Evolutionary Multimodal Reasoning via Hierarchical Semantic
Representation for Intent Recognition}

\newcommand{\correspondingauthor}[1]{
  \begingroup
  \renewcommand\thefootnote{}
  \footnotetext{#1}
  \endgroup
}

\author{Qianrui Zhou\textsuperscript{1}, Hua Xu\textsuperscript{1*}, Yunjin Gu\textsuperscript{1,3}, Yifan Wang \textsuperscript{1,2}, Songze Li \textsuperscript{1,2}, Hanlei Zhang \textsuperscript{1} \\
\textsuperscript{1}Tsinghua University \, \, \textsuperscript{2}Hebei University of Science and Technology \\
\textsuperscript{3}The Chinese University of Hong Kong
 \\
{\tt\small zgr22@mails.tsinghua.edu.cn, xuhua@tsinghua.edu.cn}
}

\begin{document}
\maketitle
\correspondingauthor{* Hua Xu is the corresponding author.}

\begin{abstract}
Multimodal intent recognition aims to infer human intents by jointly modeling various modalities, playing a pivotal role in real-world dialogue systems. However, current methods struggle to model hierarchical semantics underlying complex intents and lack the capacity for self-evolving reasoning over multimodal representations. To address these issues, we propose HIER, a novel method that integrates \textbf{HI}erarchical semantic representation with \textbf{E}volutionary \textbf{R}easoning based on Multimodal Large Language Model (MLLM). Inspired by human cognition, HIER introduces a structured reasoning paradigm that organizes multimodal semantics into three progressively abstracted levels. It starts with modality-specific tokens capturing localized semantic cues, which are then clustered via a label-guided strategy to form mid-level semantic concepts. To capture higher-order structure, inter-concept relations are selected using JS divergence scores to highlight salient dependencies across concepts. These hierarchical representations are then injected into MLLM via CoT-driven prompting, enabling step-wise reasoning. Besides, HIER utilizes a self-evolution mechanism that refines semantic representations through MLLM feedback, allowing dynamic adaptation during inference. Experiments on three challenging benchmarks show that HIER consistently outperforms state-of-the-art methods and MLLMs with 1–3\% gains across all metrics.  Code and more results are available at \url{https://github.com/thuiar/HIER}.

\end{abstract}    
\section{Introduction}
\label{sec:intro}

Multimodal intent recognition is a critical task that serves as the foundation of human-computer interaction. By utilizing diverse modalities, models gain access to broader spectrum of semantic cues, enabling comprehensive and nuanced interpretation of complex intents. Owing to its growing significance, multimodal intent recognition has been widely applied in downstream applications, including virtual humans \cite{virtualhuman2022}, intelligent transportation systems \cite{kaffash2021big}, medical diagnosis \cite{tiwari2022cnn,moon2022multimodal}, and other interactive systems \cite{paul2022intent,mi2019object}.

Although multimodal intent understanding has garnered growing interest in recent years, research in this area remains in its early stages. One core challenge lies in constructing hierarchical representations that capture the multi-level structure of complex multimodal semantics. Pioneering works \cite{zhang2022mintrec,zhang2024mintrec} have laid the foundation by releasing large-scale benchmarks spanning diverse social intents and adopting fusion strategies from multimodal sentiment analysis \cite{misa, mag-bert, mult} as baselines. More recent approaches focus on modeling fine-grained multimodal cues to enhance intent understanding, which include strengthening interactions between different modalities \cite{TCL-MAP, sdif-da, hu2025adaptive, InMu-Net, chen-etal-2024-dual-oriented}, leveraging intra- and cross-video context \cite{CAGC}, and incorporating multi-granularity optimization to improve intent comprehension \cite{zhang2024mintood}. However, these methods overlook the hierarchical nature of semantic information, which limits their ability to perform coherent and reliable reasoning. Another challenge is the reliance on suboptimal and static reasoning process, which hinders adaptive and hierarchical interpretation of multimodal semantics. Despite recent Multimodal Large Language Models (MLLMs) exhibiting strong reasoning capabilities \cite{alayrac2022flamingo,li2023blip,zhu2023minigpt,wang2024qwen2,cheng2024videollama,liu2024llavanext}, they still struggle with complex multimodal semantics due to the absence of fine-grained hierarchical reasoning path. Moreover, self-evolving refinement is crucial for dynamic and reliable reasoning, whereas existing efforts lack the capacity to generalize across complex multimodal scenarios and fail to adapt during reasoning process \cite{hosseini2024v,sun2024easy,shao2024deepseekmath,srivastava-etal-2025-debate}.



To address the above challenges, we propose HIER, a novel method which unifies hierarchical semantic modeling with self-evolving reasoning for multimodal intent recognition, as shown in Figure \ref{model}. The motivation stems from the inherently hierarchical nature of human cognition, where individuals first build situational awareness, then identify and correlate salient semantic cues, and finally synthesize them through relational reasoning reinforced by iterative self-refinement. For modeling hierarchical semantic representations, HIER begins by encoding textual and visual inputs using Qwen2-VL \cite{wang2024qwen2} encoders, forming the foundational contextual representations. To capture mid-level semantics, we propose a multimodal concept clustering strategy that groups semantically similar tokens into unified concepts via Spherical K-Means++~\cite{endo2015spherical}, where each centroid is dynamically aligned with intent semantics through cosine similarity weighting and convex combination with label vectors. For high-level relation modeling, pairwise concepts are passed through an information bottleneck network to distill compact relation whose significance is quantified via Jensen-Shannon (JS) divergence between the intent classification logits of original concepts and their combined relation. Relations with high divergence are retained for capturing novel and discriminative semantics that reflect meaningful interactions beyond individual concepts. For self-evolving reasoning, we design a structured Chain-of-Thought (CoT) that guides Qwen2-VL through three progressive stages of context understanding, concept analysis, and relation reasoning, aligned with the hierarchical semantic representation. During the concept and relation stages, detailed prompts are utilized to evaluate the plausibility of semantics, serving as the basis for self-evolution. Next, we introduce a feedback-driven mechanism for refining semantic representations through internal reasoning process. Specifically, the hidden states of both concepts and relations are projected into vocabulary logits via a shared generation head. Conditioned on prior prompts, these logits encode evaluation feedback, from which normalized confidence scores for affirmative and negative responses are derived for dynamically refining semantic features.

We summarize our contributions as follows: (1) We propose HIER, a novel framework that models hierarchical semantic representations to capture the layered nature of human cognition. To the best of our knowledge, this is the first work to establish a multi-level progressive reasoning paradigm for multimodal intent recognition. (2) We introduce a self-evolving reasoning mechanism that unifies structured CoT guidance and MLLM feedback to strengthen reasoning depth and adaptively refine semantic representations. (3) HIER achieves consistent 1–3\% gains over state-of-the-art methods and leading MLLMs on three challenging benchmarks, and generalizes effectively to enhance reasoning capabilities across diverse backbones.

\section{Related Works}
\subsection{Multimodal Intent Recognition}

Multimodal intent recognition aims to understand user intents by integrating verbal and nonverbal cues in real-world scenarios. Inspired by multimodal sentiment and emotion benchmarks \cite{zadeh2016mosi, zadeh2018mosei, yu2020ch, poria2019meld, busso2008iemocap}, MIntRec \cite{zhang2022mintrec} pioneers the task with a fine-grained dataset capturing diverse interactive intents and adopts sentiment fusion techniques \cite{misa, mag-bert, mult} as strong baselines, while MIntRec2.0 \cite{zhang2024mintrec} further advances it by expanding data scale, enriching intent taxonomy, and establishing a more comprehensive benchmark. Subsequently, specialized intent recognition methods emerge to address unique challenges in this field. For instance, TCL-MAP \cite{TCL-MAP} and MVCL-DAF \cite{hu2025adaptive} improve multimodal fusion via token-level contrastive learning, while InMu-Net \cite{InMu-Net} addresses noisy non-verbal cues using an information bottleneck and mitigates long-tail bias. Moreover, CAGC \cite{CAGC} and MuProCL \cite{dong2025unbiased} focus on the video and audio modalities respectively, capturing temporal dependencies and semantic context, thereby enhancing the understanding of non-textual modalities. To strengthen multimodal integration, SDIF-DA \cite{sdif-da} adopts a shallow-to-deep interaction framework, whereas MIntOOD \cite{zhang2024mintood} refines intent representations via weighted fusion and multi-granularity optimization. Besides, DuoDN \cite{chen-etal-2024-dual-oriented} disentangles shared and modality-specific features using a dual network with counterfactual reasoning. Recently, LGSRR \cite{zhou-etal-2025-llm} explores leveraging the reasoning capabilities of large language models to enhance intent understanding and achieves encouraging results, underscoring the pivotal role of multimodal semantic reasoning. However, its reasoning process remains relatively shallow and dependent on specific semantic concepts.


\begin{figure*}[t!]
  \centering
  \includegraphics[scale=.49]{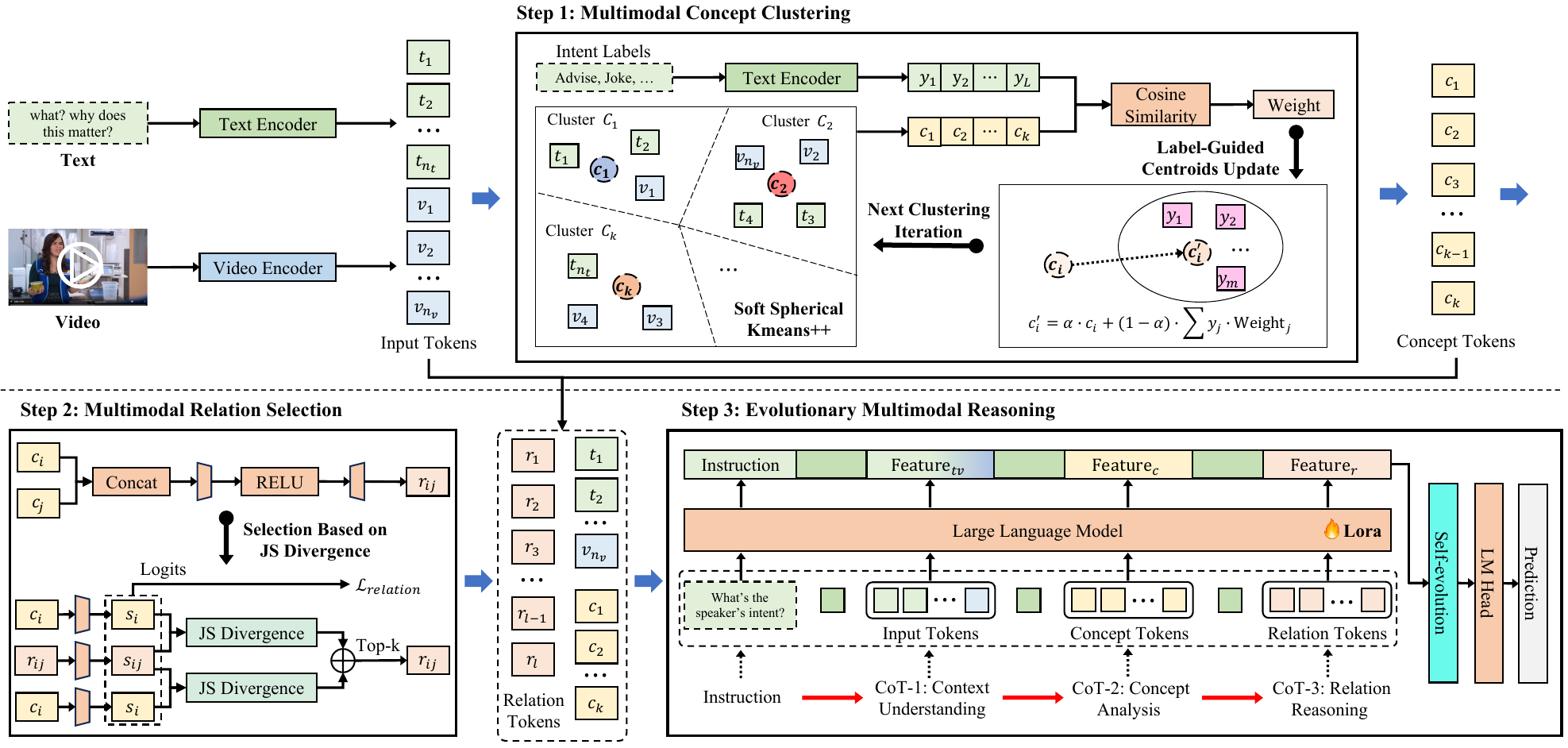}
  \caption{\label{model} Overview of the proposed HIER method. The model comprises three key steps: (1) Multimodal Concept Clustering, which groups semantically related tokens into mid-level concept representations via soft Spherical K-Means++ augmented by intent labels; (2) Multimodal Relation Selection, which captures informative inter-concept dependencies using an information bottleneck network and JS divergence; and (3) Evolutionary Multimodal Reasoning, which conducts hierarchical reasoning through a structured CoT and self-evolution mechanism, enhancing both reasoning depth and robustness.}

\end{figure*}

\subsection{Self-evolving for MLLMs}
Self-evolution mechanisms aim to enable models to autonomously acquire and refine knowledge through self-generated experiences \cite{tao2024survey,he-etal-2025-breaking,liu2025deepselfevolvingreasoning}. These strategies originate from iterative RLHF \cite{dong2024rlhf}, which aligns models with human preferences using their own generated data. Building on this foundation, self-evolution advances across various stages, including self-instruct \cite{wang-etal-2023-self-instruct}, self-play \cite{10.5555/3692070.3692326}, self-improving \cite{huang-etal-2023-large}, and self-training \cite{dou-etal-2024-rest}. Recently, the growing reasoning capabilities of Multimodal Large Language Models (MLLMs) have promoted the emergence of self-evolution approaches in multimodal scenarios. For instance, RLAIF-V \cite{yu2024rlaif} constructs preference pairs for alignment by leveraging a more capable model to evaluate the obtained responses. Similarly, SIMA \cite{wang2024enhancing} directly utilizes ground-truth answers for response comparison and preference supervision generation. Other approaches construct preference data by generating positive responses conventionally and synthesizing negative responses through deliberate perturbations, such as corrupted images \cite{zhu2024self} or misleading prompts \cite{deng2024enhancing}. Furthermore, to eliminate annotation dependence, SENA \cite{tan2024beyond} enables autonomous generation of high-quality question–answer pairs using an image-driven self-questioning mechanism. However, these methods mainly rely on preference-based alignment and lack a self-reflective reasoning process, which limits their ability to dynamically refine multimodal understanding.

\section{Method}
\subsection{Overall Architecture}
In this section, we present the architecture of HIER method, as illustrated in Figure \ref{model}. It comprises three main steps: multimodal concept clustering, multimodal relation selection, and evolutionary multimodal reasoning. The first two steps progressively extract multimodal input tokens, semantic concepts and inter-concept relations from raw data, forming a structured hierarchical representation. Building upon this, the final step performs multimodal reasoning with self-evolution, enabling both depth-aware inference and adaptive refinement for complex intent understanding.

\subsection{Multimodal Representation}

Given video data and textual input, we begin by extracting aligned modality-specific representations with Qwen2-VL encoders. The text encoder maps raw text into linguistic tokens $T = \{t_1, t_2, \dots, t_{n_t}\} \in \mathbb{R}^{n_t \times d}$, while the video encoder produces visual tokens $V = \{v_1, v_2, \dots, v_{n_v}\} \in \mathbb{R}^{n_v \times d}$, where $n_t$ and $n_v$ are the token sequence lengths for text and video, and $d$ is the unified feature dimension. Owing to the pretraining of Qwen2-VL, the extracted visual and textual tokens are embedded in a shared semantic space, ensuring semantic compatibility across modalities. These tokens are subsequently concatenated into a unified sequence $Z = \{z_1, z_2, \dots, z_{n}\} \in \mathbb{R}^{n \times d}$, where $n$ is the sum of $n_t$ and $n_v$. The obtained token sequence represents localized semantic units, forming the foundational level of hierarchical semantic modeling and serving as basic elements for the multimodal concept clustering process.

\begin{figure*}[t!]
  \centering
  \includegraphics[scale=.52]{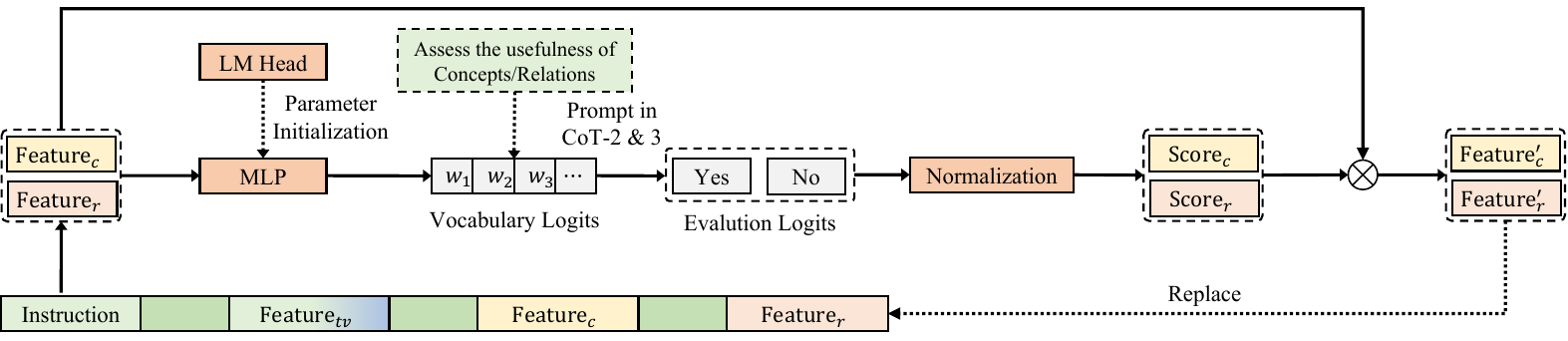}
  \caption{\label{self_evolution} Details of Self-evolution module. We first copy the Qwen2-VL’s generation head to project concept and relation features into vocabulary logits.  Conditioned on usefulness assessment prompts, we then extract and normalize logits of affirmative and negative responses to derive confidence scores, which in turn guide feature refinement for adaptive and robust reasoning.}
\end{figure*}


\subsection{Multimodal Concept Clustering}
\label{Multimodal_Concept_Clustering}

Human intent understanding hinges on intermediate cues that bridge raw multimodal signals and complex intents, narrowing the semantic gap for precise interpretation. Inspired by this cognitive process, we propose a label-guided clustering strategy to group semantically aligned tokens into mid-level concepts, enabling structured and reliable hierarchical modeling. Specifically, we adopt Spherical K-Means++ algorithm \cite{endo2015spherical} with cosine similarity, which better captures semantic proximity and mitigates magnitude bias. Given input tokens $Z$, we aim to partition them into $k$ clusters $\{C_1, \dots, C_k\}$, each represented by a centroid $c_i \in \mathbb{R}^d$ as a mid-level semantic concept. To maintain the differentiability of clustering, we utilize soft probabilistic formulations for token assignment \cite{saha2023end}. During the $u$-th iteration of clustering, we compute the probability $p_{i,m}^{(u)}$, which quantifies the contribution of token $z_i$ to the updated centroid $c_m^{(u)}$ as follows:
\begin{equation}
    p_{i,m}^{(u)} = \frac{\exp(\cos(z_i, c_m^{(u-1)}))}{\sum_{j=1}^{k}  \exp(\cos(z_i, c_j^{(u-1)}))},
\end{equation}
which is used to update centroids as $c_m^{(u)} = \sum_{i=1}^{n} p_{i,m}^{(u)} z_{i}$. 

To align clustering with intent semantics, we incorporate intent label embeddings as semantic anchors to refine centroids after each iteration. Specifically, we obtain intent vectors $\{y_1, y_2, \dots, y_L\}$ from the text encoder where $L$ is the number of labels, and compute their cosine similarities with centroids to derive alignment weight, formalized as: 
\begin{equation}
    \text{Weight}_{i,m}^{(u)} = \frac{\exp(\cos(c_m^{(u)}, y_i))}{\sum_{j=1}^{k} \exp(\cos(c_j^{(u)}, y_i))},
\end{equation}
reinforcing the coherent semantic hierarchy from localized tokens to semantic concepts and ultimately to intent-level understanding. These weights are employed to construct a label-guided semantic vector, fused with the current centroid $c_m^{(u)}$ through a convex combination:
\begin{equation}
    \tilde{c}_m^{(u)} = \alpha \cdot c_m^{(u)} + (1- \alpha) \cdot \sum_{i=1}^{L} \text{Weight}_{i,m}^{(u)} y_i,
\end{equation}
where $\alpha \in [0,1]$ balances the contribution of both parts. Through the label-guided clustering, we obtain high-quality semantic concepts $\{c_1, c_2, \dots, c_k\}$ that capture salient mid-level patterns essential for intent comprehension.

\subsection{Multimodal Relation Selection}
\label{Multimodal_Relation_Selection}

To elevate mid-level semantic concepts for high-level reasoning, we model their interactions via a relation selection strategy that captures informative inter-concept dependencies. Specifically, we employ an information bottleneck network \cite{Federici2020Learning} to encode compact relational representations and use Jensen-Shannon (JS) divergence to quantify the semantic novelty of each relation as it extends Kullback–Leibler (KL) divergence with improved balance and symmetry. Crucially, high-divergence relations capture complementary and emergent semantics beyond individual concepts, aligning with the human reasoning process where semantic information arises not only from individual concepts but also from their interactions, thereby enriching the depth and coherence of intent understanding.



Given concept representations $\{c_1, c_2, \dots, c_k\}$, we consider all pairwise combinations $(c_i, c_j)$ and encode their interactions $r_{ij}$ using an information bottleneck network which consists of non-linear layers with ReLU activation:
\begin{equation}
r_{ij} = \text{MLP}(\text{ReLU}([c_i;c_j])),
\end{equation}
where $[c_i;c_j]$ denotes the concatenation of concept vectors.
To ensure alignment with intent semantics, both concepts $(c_i, c_j)$ and their relations $r_{ij}$ are jointly optimized using intent classification objectives:
\begin{equation}
s_{i} = \text{MLP}_c(c_i), \, s_{j} = \text{MLP}_c(c_j),
\end{equation}
\begin{equation}
s_{ij} = \text{MLP}_r(r_{ij}),
\end{equation}
where $\text{MLP}_c(\cdot)$ and $\text{MLP}_r(\cdot)$ denote classification heads applied to concept and relation representations, respectively. Furthermore, $\mathcal{L}_{\text{relation}}$ is defined as:
\begin{equation}
    \mathcal{L}_{\text{relation}} = \text{CE}(s_i, g) + \text{CE}(s_j, g) + \text{CE}(s_{ij}, g),
\end{equation}
\begin{equation}
    \text{CE}(s, g) = -\frac{1}{B} \sum_{i=1}^{B} \sum_{j=0}^{|\mathcal{Y}|} \mathbb{I}_{[g_i = j]} \log \left( \operatorname{Softmax}(s_i)^j \right),
\end{equation}
where $g$ represents the ground truth intent, $B$ is the batch size, and $\mathcal{Y} = \{0, 1, \cdots, L-1\}$ denotes the label set. Next, we assess the semantic gain from combining $c_i$ and $c_j$ by computing the JS divergence between the relation logits $s_{ij}$ and individual concept logit. Taking $\text{JS}_i$ as an example: 
\begin{equation}
    m_i = \frac{1}{2}(\text{Softmax}(s_i) + \text{Softmax}(s_{ij})),
\end{equation}
\begin{equation}
    \text{JS}_{i} = \frac{1}{2} (\text{KL}(\text{Softmax}(s_i) \| m_i) + \text{KL}(m_i \| \text{Softmax}(s_{ij}))),
\end{equation}
where $\text{KL}(\cdot | \cdot)$ denotes the calculation of KL divergence. We compute the final score as $\text{JS}_{ij} = \text{JS}_i + \text{JS}_j$, and select the top-$k$ relations with the highest scores. Given that the scale of divergence may differ across instances, we employ a fixed retention ratio to adaptively preserve the top-ranking relations within each sample, ensuring effective selection of informative relation representations $\{r_1, r_2, \dots, r_l\}$.

\subsection{Evolutionary Multimodal Reasoning}
\label{Evolutionary_Multimodal_Reasoning}

Building on the hierarchical semantics, we guide Qwen2-VL to perform progressive multimodal reasoning using a structured Chain-of-Thought (CoT) prompt that mirrors human cognition, progressing through context perception, concept abstraction, and relational analysis. The CoT-driven multimodal reasoning unfolds through three progressive stages that align with different semantic levels. In CoT-1 (Context Understanding), modality-specific input tokens $Z = \{z_1, z_2, \dots, z_n\}$ are first processed to establish a global understanding of the multimodal scene. CoT-2 (Concept Analysis) then focuses on mid-level semantic concepts, using tokens $\{c_1, c_2, \dots, c_k\}$ to distill key intent-relevant cues from the context. Finally, CoT-3 (Relational Reasoning) introduces relation tokens $\{r_1, r_2, \dots, r_l\}$ to model interactions between concepts, capturing higher-order semantic dependencies crucial for nuanced intent inference. Notably, we explicitly prompt the model with \textit{judge their usefulness} in the final two stages to encourage reflection on concepts and relations, laying the foundation for self-evolution. Besides, hierarchical semantic features are initially denoted by special tokens (e.g., \textless concept\textgreater~and \textless relation\textgreater) and then seamlessly integrated into the encoded prompt, jointly guiding Qwen2-VL through coherent and in-depth reasoning. 
Appendix 2.1 and 4.1 present detailed prompt design and sensitivity analyses, and Appendix 4.2 provides comparison with CoT-enhanced MLLMs.


To enhance adaptability in complex scenarios, we introduce a self-evolution mechanism that enables internal evaluation and refinement of semantic features, as illustrated in Figure \ref{self_evolution}. After Qwen2-VL processes the prompt representation, we isolate concept and relation features ($\text{Feature}_c$ and $\text{Feature}_r$), and project them into vocabulary logits ($\text{Logits}_c$ and $\text{Logits}_r$) using a linear layer initialized with Qwen2-VL’s generation head for alignment:
\begin{equation}
\text{Logits}_c = \text{MLP}_{lm}(\text{Feature}_c), \, 
\end{equation}
\begin{equation}
    \text{Logits}_r = \text{MLP}_{lm}(\text{Feature}_r),
\end{equation}
where $\text{MLP}_{lm}(\cdot)$ refers to the shared linear layer. Leveraging the reflection prompts in CoT-2 and CoT-3 as contextual cues, we directly extract the logits of affirmative and negative responses (e.g., “Yes” and “No”) and normalize them into refinement scores for the self-evolution process:
\begin{equation}
\text{Score}_c = \frac{\exp(\text{Logits}_c[Idx^+])}{\exp(\text{Logits}_c[Idx^+]) + \exp(\text{Logits}_c[Idx^-])},
\end{equation}
\begin{equation}
\text{Score}_r = \frac{\exp(\text{Logits}_r[Idx^+])}{\exp(\text{Logits}_r[Idx^+]) + \exp(\text{Logits}_r[Idx^-])},
\end{equation}
where $Idx^+$ and $Idx^-$ denote the logits corresponding to affirmative and negative responses, respectively. The refinement scores are used to modulate each semantic feature by its corresponding confidence score, enhancing reasoning robustness and adaptability. Finally, the refined features is computed as $\text{Feature}' = \text{Score} \cdot \text{Feature}$ and replace the original representations in the reasoning-aware features.

The self-evolved features are fed into the generation head to compute the autoregressive language modeling loss \cite{wang2024qwen2} $\mathcal{L}_{\text{task}}$. The overall training objective combines this task-specific loss with the relation-guided auxiliary loss:
\begin{equation}
    \mathcal{L}= \mathcal{L}_{\text{task}} + \beta \mathcal{L}_{\text{relation}},
\end{equation}
where $\beta$ balances the contributions of both losses.

\begin{table}[t!]
    \fontsize{8.85}{10.8}\selectfont %
\setlength{\tabcolsep}{5pt} %
\caption{Statistics of MIntRec, MIntRec2.0, and MELD-DA datasets. \#C denote the number of classes.}
  \label{dataset_statistics}
\centering
  \begin{tabular}{lcccc}
    \toprule
    \textbf{Datasets} & \textbf{\#C} & \textbf{\#Train} & \textbf{\#Validation} & \textbf{\#Test} \\
    \midrule
    MIntRec & 20 & 1,334 & 445 & 445 \\
    MIntRec2.0 (in-scope) & 30 & 6,165 & 1,106 & 2,003 \\
    MELD-DA & 12 & 6,991 & 999 & 1,998 \\
    \bottomrule
  \end{tabular}
\end{table}

\begin{table*}[t!]
\fontsize{8.65}{10.8}\selectfont 
\setlength{\tabcolsep}{2.25pt} 
\caption{Comparison results with SOTA methods and leading MLLMs, with best scores in bold and second-best underlined.}
\label{main_results}
\centering
\renewcommand{\arraystretch}{1.4}

\begin{tabular}{l|cccccc|cccccc|cccccc}
\toprule
Datasets & 
\multicolumn{6}{c|}{MIntRec} & 
\multicolumn{6}{c|}{MIntRec2.0} & 
\multicolumn{6}{c}{MELD-DA} \\
\midrule
 Methods & ACC & F1 & P & R & WF1 & WP & ACC & F1 & P & R & WF1 & WP & ACC & F1 & P & R & WF1 & WP \\
\midrule

MISA & 72.29 & 69.24 & \underline{72.38} & \underline{73.48} & 69.32 & 70.85 & 55.16 & 49.51 & 51.80 & 49.92 & 55.05 & 57.06 & 60.86 & 49.45 & 52.70 & 49.14 & 58.80 & 59.55 \\
MAG-BERT & 72.40 & 68.29 & 68.87 & 69.22 & 72.06 & 72.94 & 60.38 & \underline{54.74} & 57.51 & \underline{54.54} & \underline{59.61} & 60.00 & 61.08 & 50.02 & 52.29 & 49.85 & 59.59 & 59.60 \\
MulT & 72.31 & 68.97 & 69.73 & 68.83 & 72.07 & 72.24 & \underline{60.66} & 54.12 & \underline{58.02} & 53.77 & 59.55 & \underline{60.12} & 59.99 & 50.69 & 54.83 & 50.59 & 58.67 & 59.39 \\
TCL-MAP & 73.17 & 68.92 & 68.90 & 69.99 & 72.66 & 72.97 & 58.24 & 52.25 & 54.28 & 52.41 & 57.24 & 57.55 & \underline{61.63} & 50.25 & 53.32 & 49.64 & \underline{59.74} & 60.16 \\
SDIF-DA & 71.64 & 68.19 & 69.08 & 68.30 & 71.34 & 71.74 & 58.06 & 51.95 & 53.17 & 52.16 & 57.47 & 57.85 & 60.91 & \underline{51.46} & \underline{57.57} & \underline{50.80} & 59.58 & 60.17 \\
MIntOOD & 73.48 & \underline{70.51} & 71.19 & 70.45 & 72.39 & 73.24 & 58.73 & 52.40 & 55.48 & 51.20 & 58.03 & 58.34 & 61.58 & 50.57 & 54.97 & 50.63 & 59.46 & \underline{60.37} \\
MVCL-DAF & \underline{73.63} & 70.41 & 71.07 & 70.11 & \underline{73.57} & \underline{74.31} & 59.64 & 53.41 & 54.90 & 53.24 & 58.67 & 58.57 & 60.78 & 48.88 & 54.05 & 47.73 & 59.16 & 59.83 \\
\midrule
Qwen2-VL & \underline{76.56} & \underline{74.59} & \underline{75.85} & \underline{74.76} & \underline{76.44} & \underline{77.19} & 59.82 & 47.73 & \underline{55.23} & 47.25 & 58.44 & \underline{62.51} & 59.71 & \underline{52.44} & 55.19 & \underline{51.93} & 59.11 & 59.15 \\
LLaVA-NeXT & 72.65 & 64.94 & 66.21 & 64.65 & 72.63 & 73.59 & 50.61 & 45.33 & 50.79 & 44.14 & 50.70 & 54.21 & 51.30 & 39.87 & 43.68 & 38.36 & 49.77 & 50.52 \\
VideoLLaMA2 & 74.61 & 71.64 & 71.82 & 73.35 & 74.37 & 75.43 & \underline{60.11} & \underline{49.95} & 51.58 & \underline{49.13} & \underline{59.71} & 59.90 & \underline{60.81} & 51.24 & \underline{56.20} & 48.79 & \underline{59.53} & \underline{59.79} \\
\midrule
HIER & \textbf{80.00} & \textbf{76.91} & \textbf{78.98} & \textbf{77.11} & \textbf{79.59} & \textbf{80.67} & \textbf{64.15} & \textbf{60.31} & \textbf{61.90} & \textbf{59.59} & \textbf{63.79} & \textbf{64.17} & \textbf{61.95} & \textbf{54.80} & \textbf{59.41} & \textbf{52.94} & \textbf{60.38} & \textbf{60.44} \\

\bottomrule
\end{tabular}
\end{table*}

\section{Experiment} 

\subsection{Datasets and Evaluation Metrics} 
We use MIntRec \cite{zhang2022mintrec} and the in-scope subset of MIntRec2.0 \cite{zhang2024mintrec} for multimodal intent recognition, and MELD-DA \cite{EMoTyDA} for multimodal dialogue classification. Detailed statistics are shown in Table \ref{dataset_statistics}. Following \cite{zhang2022mintrec, zhang2024mintrec}, we adopt classification metrics for evaluation, including accuracy (ACC), F1-score (F1), precision (P), recall (R), weighted F1-score (WF1), and weighted precision (WP).

\subsection{Baselines}
We compare HIER against state-of-the-art methods and leading MLLMs, including: (1) \textbf{MISA} \cite{misa}, \textbf{MAG-BERT} \cite{mag-bert}, and \textbf{MulT} \cite{mult} are adapted from multimodal sentiment analysis; (2) \textbf{TCL-MAP} \cite{TCL-MAP} enhances textual cues via modality-aware prompting, label-based augmentation, and token-level contrastive alignment; (3) \textbf{SDIF-DA} \cite{sdif-da} refines nonverbal alignment through a shallow-to-deep interaction module; (4) \textbf{MIntOOD} \cite{zhang2024mintood} combines weighted fusion and multi-granularity modeling for robust representation learning; (5) \textbf{MVCL-DAF} \cite{hu2025adaptive} employs multi-view contrastive learning with dynamic attention allocation fusion; and (6) \textbf{Qwen2-VL} \cite{wang2024qwen2}, \textbf{VideoLLaMA2} \cite{cheng2024videollama}, and \textbf{LLaVA-NEXT} \cite{liu2024llavanext} are strong representatives of recent advances in general-purpose multimodal reasoning.

\subsection{Experimental Settings}

We build HIER upon Qwen2-VL and fine-tune it using LoRA \cite{hu2022lora} with a rank of 8 and a batch size of 2, implemented via LLaMA-Factory \cite{zheng2024llamafactory}. Optimization is performed using AdamW \cite{loshchilov2017decoupled} optimizer, with loss balancing coefficient $\beta$ set to 0.01. The input sequence length is dynamically adapted following Qwen2-VL’s configuration, with the hidden dimension set to 3,584. The clustering process runs for 30 iterations, and the label-guided interpolation coefficient $\alpha$ is learned during training. For MIntRec, we use 50 concept and 25 relation tokens, while for MIntRec2.0 and MELD-DA, these are set to 40 and 20, respectively. Following prior work, we report results averaged over five random seeds 0–4. For our method and MLLMs, fine-tuning is conducted for 5 epochs using an NVIDIA A100-PCIE GPU. Detailed training cost analysis and hyper parameters are included in Appendix 3.1 and 3.2.

\subsection{Main Results}

Table~\ref{main_results} presents the main results on three challenging datasets, with best scores in bold and second-best underlined. Compared to multimodal intent recognition baselines, HIER achieves 3–10\% gains across all metrics on both MIntRec and MIntRec2.0. While TCL-MAP and SDIF-DA remain competitive on MELD-DA, HIER still yields over 2\% gains on half of the metrics. The improvements are particularly notable on datasets with fine-grained intents, highlighting HIER’s strength in capturing complex hierarchical semantics and supporting progressive intent reasoning. Furthermore, HIER delivers the highest F1 scores on all three datasets, with significant margins of 7.67\%, 5.57\%, and 3.34\%, reflecting its robustness across diverse intent classes and ability to interpret complex semantics driven by the evolutionary reasoning paradigm. In comparison with leading MLLMs, HIER consistently achieves superior performance across all three datasets. On MIntRec, it outperforms the strong Qwen2-VL baseline by over 2\% on all metrics, while on the more challenging MIntRec2.0, it exhibits absolute advantages on  F1, P, and R metrics with improvements of 10.36\%, 6.67\%, and 10.46\%, respectively. Although VideoLLaMA2 performs competitively on MELD-DA, especially on WF1 and P metrics, HIER still maintains a clear performance lead with gains exceeding 0.5\%, further validating its superiority across diverse scenarios. Despite the impressive semantic reasoning capabilities of MLLMs, HIER surpasses them by harnessing explicitly structured semantic hierarchies and self-evolving reasoning mechanism, closely aligned with human cognitive processes for multimodal intent understanding. Additional performance variability and significance test are provided in Appendix 3.3, while a detailed case study is presented in Appendix 4.5.

\begin{table}[t!] 
\fontsize{8.85}{10.8}\selectfont 
\setlength{\tabcolsep}{2.7pt} 
\caption{Ablation results of different modules on the MIntRec, MIntRec2.0, and MELD-DA datasets.}
\label{ablation_results}
\centering
\renewcommand{\arraystretch}{1.1} 

\begin{tabular}{cl|cccccc}
\toprule
 & Ablations & ACC & F1 & P & R & WF1 & WP \\
\midrule

\multirow{5}{*}{\rotatebox{90}{MIntRec}} 
    & w/o Concept          & \underline{79.55} & \underline{76.11} & \underline{78.06} & \underline{76.31} & \underline{79.25} & \underline{80.30} \\
    & w/o Relation         & 77.08 & 70.76 & 71.55 & 71.52 & 76.92 & 77.83 \\
    & w/o CoT              & 75.51 & 71.57 & 72.03 & 72.36 & 74.95 & 75.37 \\
    & w/o Self-evolution  & 77.75 & 71.50 & 72.05 & 72.31 & 77.53 & 78.20 \\
    \cmidrule{2-8}
    & HIER                 & \textbf{80.00} & \textbf{76.91} & \textbf{78.98} & \textbf{77.11} & \textbf{79.59} & \textbf{80.67} \\
\midrule\midrule
\multirow{5}{*}{\rotatebox{90}{MIntRec2.0}} 
    & w/o Concept          & 63.07 & 50.13 & 50.39 & 50.44 & 62.80 & 62.99 \\
    & w/o Relation         & 62.67 & 55.73 & 57.18 & 56.19 & \underline{62.95} & \underline{64.09} \\
    & w/o CoT              & \underline{63.35} & 56.81 & 58.59 & 56.19 & 62.87 & 63.27 \\
    & w/o Self-evolution  & 63.06 & \underline{58.38} & \underline{60.61} & \underline{57.11} & 62.39 & 62.88 \\
    \cmidrule{2-8}
    & HIER                 & \textbf{64.15} & \textbf{60.31} & \textbf{61.90} & \textbf{59.59} & \textbf{63.79} & \textbf{64.17} \\
\midrule\midrule

\multirow{5}{*}{\rotatebox{90}{MELD-DA}} 
    & w/o Concept          & 59.90 & 51.85 & 55.24 & \underline{51.13} & \underline{59.07} & \underline{59.25} \\
    & w/o Relation         & 60.01 & 51.85 & 55.84 & 50.14 & 58.76 & 58.49 \\
    & w/o CoT              & 59.91 & \underline{52.08} & \underline{56.73} & 50.13 & 58.62 & 58.46 \\
    & w/o Self-evolution  & \underline{60.21} & 50.50 & 53.72 & 48.90 & 58.74 & 58.53 \\
    \cmidrule{2-8}
    & HIER                 & \textbf{61.95} & \textbf{54.80} & \textbf{59.41} & \textbf{52.94} & \textbf{60.38} & \textbf{60.44} \\
\bottomrule
\end{tabular}
\end{table}


\subsection{Ablation Study}
To assess the contribution of each module in HIER, we conduct ablations under four settings: w/o Concept, w/o Relation, w/o CoT, and w/o Self-evolution, as shown in Table \ref{ablation_results}. \textbf{(1) w/o Concept:} To assess the contribution of semantic concepts, we remove the multimodal concept clustering step and directly model relations among modality-specific tokens. This leads to a notable drop exceeding 1\% across all metrics on MIntRec2.0 and MELD-DA, highlighting the importance of structured concept representation. While the performance on MIntRec remains comparable, likely due to its smaller scale limiting mid-level semantics learning, a consistent decline of over 0.3\% across all metrics is still observed. \textbf{(2) w/o Relation:} Disabling the relation selection module causes significant degradation across most metrics, with drops 1-7\% on all datasets, except for WP on MIntRec2.0. This outcome aligns with prior works \cite{TCL-MAP,sdif-da} emphasizing the critical role of modeling cross-modal dependencies and confirms the effectiveness of our selection strategy in identifying semantically informative inter-concept relations. \textbf{(3) w/o CoT:} To demonstrate the benefit of the progressive reasoning in leveraging hierarchical semantics, we replace the CoT prompts with straightforward instructions. Notable performance drops are observed across all datasets, especially on F1, P, and R metrics with declines exceeding 2\%, revealing the effectiveness of the shallow-to-deep reasoning trajectory in enhancing complex intent understanding through hierarchical semantic representations. \textbf{(4) w/o Self-evolution:} Removing the self-evaluation module results in the largest performance drop, especially on MELD-DA with a 5.69\% decline in P metric, confirming its role in refining semantic representations and improving reasoning robustness. The absence also leads to consistent drops of 2–6\% on MIntRec and 1–2\% on MIntRec2.0, highlighting the importance of the self-evolution mechanism for overall performance.

\subsection {Impact of Concept and Relation Quantity}
\begin{figure}[t!]
  \centering
  \includegraphics[width=\linewidth]{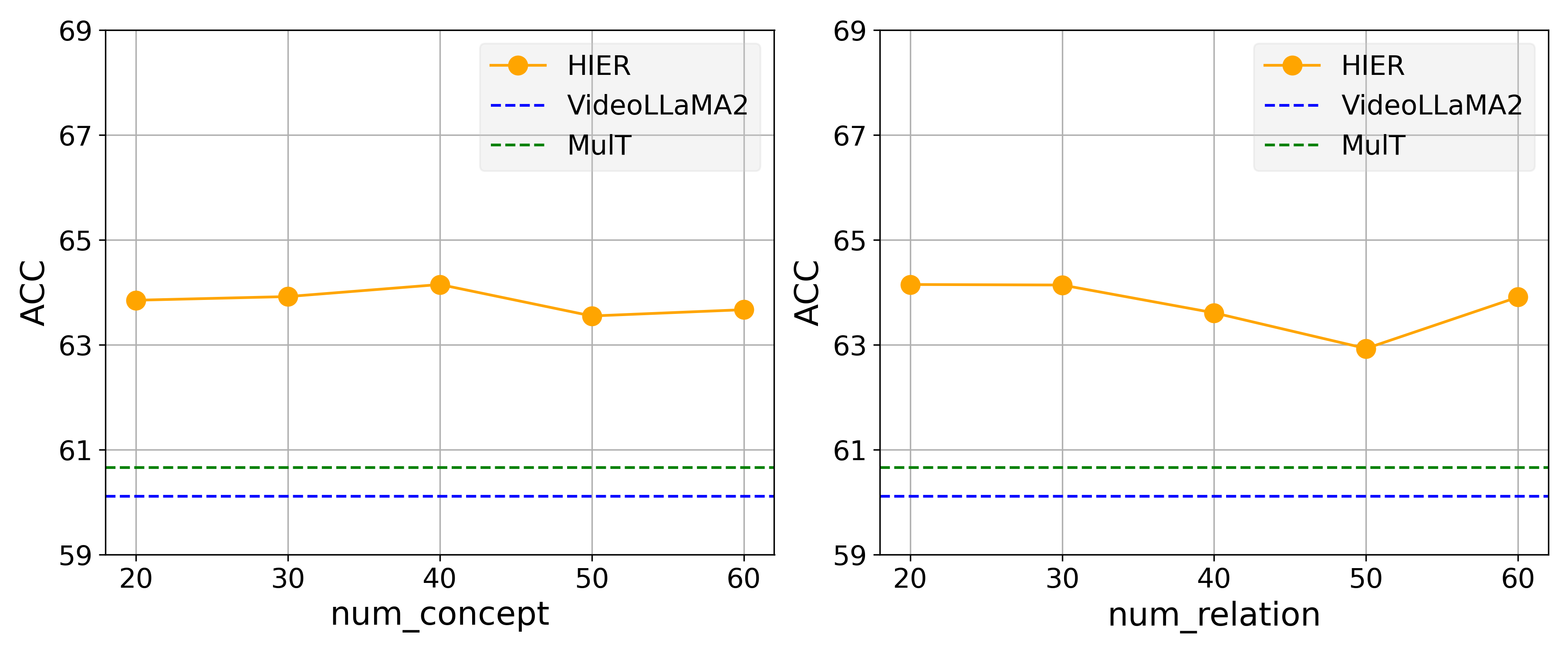}
  \caption{Impact of concept and relation quantity in HIER evaluated on the MIntRec2.0 dataset.}
  \label{concept_relation_quantity}
\end{figure}

To examine the effect of concept and relation quantity, we evaluate HIER on MIntRec2.0 with ACC, comparing to the strongest baselines including MulT and VideoLLaMA2, as shown in Figure \ref{concept_relation_quantity}. First, we vary concept tokens from 20 to 60 in steps of 10, maintaining a fixed 2:1 ratio between concepts and relations to keep relation pairs within a reasonable range. Next, with concept tokens fixed at 40, we vary relation tokens from 20 to 60 to assess their independent impact. These settings allow for a systematic analysis of how the depth and granularity of the hierarchical semantic structure influence overall model accuracy. As the number of concept tokens increases, HIER consistently sustains strong performance and achieves a peak ACC of 64.15\% at 40 concepts, substantially outperforming the baselines, which confirms the effectiveness of our label-guided clustering in robustly discovering intent-relevant semantic concepts, regardless of variations in the target number of clusters. For relation tokens, HIER shows remarkable stability with only a slight dip to 62.93\% on ACC at 50 tokens, still exceeding both baselines by over 1\%. This stability underscores the strength of our multimodal relation selection strategy in distilling semantically rich and intent-discriminative relations. By effectively filtering out redundant or noisy correlations, the model preserves only those relational cues that truly matter for understanding complex intents, reinforcing the value of structured reasoning in multimodal contexts. 


\begin{table}[t!]
\fontsize{9}{10.8}\selectfont 
\setlength{\tabcolsep}{4pt} 
\caption{Performance of HIER with different MLLM backbones on the MIntRec datastet.}
\label{backbone_experiment}
\centering
\renewcommand{\arraystretch}{1.1} 

\begin{tabular}{l|cccccc}
\toprule
Methods & ACC & F1 & P & R & WF1 & WP \\
\midrule
LLaVA-NeXT        & 72.65 & 64.94 & 66.21 & 64.65 & 72.63 & 73.59 \\
\multicolumn{1}{r|}{\textbf{\textit{$\text{+HIER}$}}}  & \textbf{78.20} & \textbf{75.33} & \textbf{75.32} & \textbf{76.30} & \textbf{77.97} & \textbf{78.59} \\
VideoLLaMA2       & 74.61 & 71.64 & 71.82 & 73.35 & 74.37 & 75.43 \\
\multicolumn{1}{r|}{\textbf{\textit{$\text{+HIER}$}}} & \textbf{77.18} & \textbf{74.56} & \textbf{75.33} & \textbf{75.06} & \textbf{76.35} & \textbf{77.27} \\
\bottomrule
\end{tabular}

\end{table}

\begin{table*}[t!]
\fontsize{9}{10.8}\selectfont 
\setlength{\tabcolsep}{3pt} 
\caption{\label{results_each_classes} F1-score comparison between HIER and outperforming baselines for each class on the MIntRec dataset.
}
\centering
\renewcommand{\arraystretch}{0.9} 
\begin{tabular*}{0.99\textwidth}{c|c@{\extracolsep{\fill}}cccccccccccc} 
 \toprule
\multicolumn{1}{l|}{Methods} &Complain& Praise & Apologise & Thank & Criticize  & Care & Agree & Taunt & Flaunt & Oppose & Joke   \\ 
\midrule
  \multicolumn{1}{l|}{MISA}  & 63.91 & 86.63 & 97.78 & \textbf{98.03} & 53.44 & 87.14 & 92.05 & 22.15 & 46.44 & 36.15 & 38.74 \\
  \multicolumn{1}{l|}{TCL-MAP}  & 68.70 & 87.20 & 97.70 & 97.00 & 51.30 & 86.80 & \underline{93.10} & 17.20 & 50.80 & 35.90 & 29.00 \\
  \multicolumn{1}{l|}{VideoLLaMA2} & \underline{75.00} & \textbf{93.02} & \underline{98.18} & 92.00 & \underline{63.16} & 82.93 & \textbf{96.00} & \textbf{35.29} & \textbf{63.64} & 50.00 & 47.62 \\
 \multicolumn{1}{l|}{Qwen2-VL } & 74.14 & 88.10 & 94.55 & 95.83 & 56.00 & \underline{89.47} & 77.42 & \underline{31.58} & \underline{63.16} & \textbf{80.00} & \underline{63.16}  \\
 \midrule
\multicolumn{1}{l|}{HIER}  & \textbf{79.67} & \underline{88.37} & \textbf{100.00} & \underline{97.96} & \textbf{69.57} & \textbf{94.74} & 85.71 & 23.53 & 60.87 & \underline{66.67} & \textbf{66.67} \\
  \multicolumn{1}{l|}{Human}  & 80.08 & 93.44 & 96.15 & 96.90 & 72.21 & 96.09 & 87.21 & 65.55 & 78.10 & 69.04 & 72.22 \\
\end{tabular*}

\begin{tabular*}{0.99\textwidth}{c|c@{\extracolsep{\fill}}cccccccccc} 
 \midrule
  \midrule
\multicolumn{1}{l|}{Methods} & Inform  &Advise & Arrange & Introduce & Comfort & Leave & Prevent & Greet & Ask for help\\
\midrule
  \multicolumn{1}{l|}{MISA}  & 70.18 & 69.56 & \underline{67.32} & 67.22 & 78.78 & \underline{77.23} & \underline{83.30} & 82.71 & \underline{67.57} \\
\multicolumn{1}{l|}{TCL-MAP}  & \underline{72.80} & 68.90 & 65.40 & 68.40 & 79.80& \textbf{83.40} & \textbf{83.60} & \textbf{90.10} & 66.40 \\
  \multicolumn{1}{l|}{VideoLLaMA2}  & 62.86 & \textbf{81.48} & 51.06 & 71.79 & \underline{88.89} & 64.86 & 75.86 & \underline{84.62} & 54.55 \\
 \multicolumn{1}{l|}{Qwen2-VL } & 70.27 & \underline{79.17} & 66.67 & \textbf{85.00} & 84.85 & 68.57 & 81.25 & 80.00 & 62.50 \\
\midrule
\multicolumn{1}{l|}{HIER}  & \textbf{75.47} & 78.26 & \textbf{72.73} & \underline{76.19} & \textbf{91.89} & 70.59 & 81.48 & 80.00 & \textbf{77.78} \\
  \multicolumn{1}{l|}{Human}  & 79.69 & 87.14 & 81.40 & 84.09 & 95.95 & 97.06 & 86.43 & 94.15 & 88.54 \\
\bottomrule
 \end{tabular*}
\end{table*}

\subsection{Performance with Different MLLM Backbones}

To further assess generalizability of HIER across different backbones, we integrate it with the other two representative MLLMs, LLaVA-NeXT and VideoLLaMA2, employing the same hyper parameters and experimental settings as in the main experiments. As shown in Table \ref{backbone_experiment}, integrating HIER with LLaVA-NeXT yields consistently remarkable improvements, surpassing 5\% across all metrics. Particularly striking are the gains in F1 and R metrics which exceed 10\%, underscoring that HIER substantially enhances the reasoning ability of backbones by not only employing hierarchical semantic modeling to deepen and structure reasoning process for greater depth and logical coherence, but also utilizing the self-evolution mechanism to continuously refine intermediate representations for reasoning robustness. When applied to VideoLLaMA2, HIER also delivers remarkable gains elevating every metric but F1 beyond 75\%, whereas the original model exceeded this threshold only on WP metric. These results demonstrate HIER’s strong generalizability across backbones, effectively enhancing existing MLLMs’ reasoning over complex multimodal semantics. Results with latest backbones and comparisons with closed-source models are shown in Appendix 4.3 and 4.4.

\subsection{Performance on Fine-Grained Intent Classes}

To provide deeper insight into the effectiveness of our method, we report the F1-scores for each fine-grained intent label on MIntRec, as shown in Table \ref{results_each_classes}. The comparison includes the top two multimodal intent recognition methods (TCL-MAP and MISA) and two leading MLLMs (Qwen2-VL and VideoLLaMA2), selected based on overall F1 performance. Furthermore, human evaluation from \cite{zhang2022mintrec} is utilized to identify potential improvements, offering a valuable reference for uncovering model limitations. From the overall perspective, HIER demonstrates a dominant advantage in fine-grained intent recognition, achieving the highest F1 scores in 9 out of 20 categories, nearly double the second-best results of 5 attained by VideoLLaMA2. This clear margin demonstrates HIER’s superior capacity and robustness in handling the majority of nuanced intent categories. Specifically, in 5 out of these 9 intent categories, our method achieves a substantial improvement of more than 4\% over the second-best scores, highlighting its effectiveness in capturing complex and discriminative intent semantics. Besides, HIER achieves a remarkable 100\% F1 score on the \textit{Apologise} intent category, serving as compelling evidence for the effectiveness of our proposed hierarchical representations and evolutionary reasoning method. Compared to human evaluation, HIER delivers remarkably competitive results, outperforming human performance on the \textit{Apologise} and \textit{Thank} categories, and achieving a margin of less than 2\% in three other categories. This strong performance showcases the significance of our reasoning paradigm, which emulates the stepwise and hierarchical nature of human intent comprehension. However, in the \textit{Taunt}, \textit{Flaunt}, and \textit{Leave} categories, HIER falls noticeably short of human performance, likely due to the semantic inconsistencies, such as mismatched facial expressions and spoken content. These challenges point to critical areas for future exploration, particularly in enhancing the model’s capacity to resolve conflicts in complex scenarios.


\section{Conclusion}

In this paper, we introduce HIER, a novel method that unifies hierarchical semantic representation with self-evolving reasoning to tackle the challenges of capturing progressive semantics and conducting self-evolving reasoning for multimodal intent recognition. HIER constructs a layered semantic hierarchy, ranging from modality-specific tokens, to mid-level intent-aware concepts, and finally to inter-concept relations. This is achieved by a label-guided clustering strategy that forms semantically aligned concept representations and an information-theoretic mechanism that effectively selects discriminative inter-concept relations. Based on this hierarchy, HIER performs cognitively inspired and progressive reasoning through a structured CoT strategy, further enhanced by a self-evolution module that refines semantic features via internal feedback. Experiments on three challenging datasets showcase HIER’s superior performance over SOTA baselines and leading MLLMs, which collectively affirms its pioneering role in establishing a novel evolutionary multimodal reasoning paradigm.

\section*{Acknowledgement}
This paper is funded by National Natural Science Foundation of China (Grant No. 62173195).
{
    \small
    \bibliographystyle{ieeenat_fullname}
    \bibliography{main}
}


\end{document}